\documentclass[12pt]{article}
\usepackage{amssymb}\usepackage{amsmath}\usepackage{amsthm}

\def \RR {{\mathbb R}}
\def \CC {{\mathbb C}}

\def \NN {{\mathbb N}}

\def\bea{\begin{eqnarray}} 
\def\ba{\begin{array}} \def\eea{\end{eqnarray}} 
\def\ea{\end{array}}

\def\ovl{\overline}
\def\d{\partial}

\newcommand{\qbox}[1]{\quad\hbox{#1}\quad}
\newcommand{\qqbox}[1]{\qquad\hbox{#1}\qquad}
\newcommand{\supp}{{\rm supp\>}}
\newcommand{\beq}{\begin{equation}}
\newcommand{\eeq}{\end{equation}}
\newtheorem{prop}{Proposition}
\newtheorem{thm}[prop]{Theorem}

\begin{document}

\title{Connection between the renormalization groups of St\"uckelberg-Petermann
and Wilson}


\author{Michael D\"utsch$^{1,2}$}

\maketitle

\vskip-10mm

\begin{center}
\footnotesize
$^1$ Institut f\"ur Theoretische Physik, Universit\"at G\"ottingen, \\
Friedrich-Hund-Platz 1, 37077 G\"ottingen, Germany \\[2mm]
$^2$ Courant Research Centre ``Higher Order Structures in Mathematics'',\\
  Universit\"at G\"ottingen, Bunsenstr.\ 3--5, 37073 G\"ottingen, Germany \\[2mm]
michael.duetsch@theorie.physik.uni-goe.de\\
\end{center}

\vskip10mm

\begin{abstract}
The St\"uckelberg-Petermann renormalization group is the group of finite renormalizations of the $S$-matrix in the 
framework of causal perturbation theory. The renormalization group in the sense of Wilson relies usually on 
a functional integral formalism, it describes the
dependence of the theory on a UV-cutoff $\Lambda$; a widespread procedure is to construct the theory by solving 
Polchinski's flow equation for the effective potential.

To clarify the connection between these different approaches we proceed as follows: in 
the framework of causal perturbation theory we introduce an UV-cutoff $\Lambda$, 
define an effective potential $V_\Lambda$, prove a
pertinent flow equation and compare with the corresponding terms in the functional integral formalism.
The flow of $V_\Lambda$ is a version of Wilson's renormalization group. The restriction of these operators to 
local interactions can be approximated by a subfamily of the  St\"uckelberg-Petermann renormalization group.
\end{abstract}

\section{Introduction}\setcounter{equation}{0}

There are different versions of the renormalization group (RG), however their relations are not 
completely understood.
The aim of this article is to clarify the connection between the renormalization groups of 
St\"uckelberg-Petermann and Wilson in the framework of perturbation theory. 
In order that this paper is better intelligible for physicists we use sometimes 
a graphical language and omit some mathematical technicalities, for a more mathematical 
formulation we refer to \cite{BDF}.
 
By the St\"uckelberg-Petermann RG and Wilson's RG we mean the following.
\begin{itemize}
\item The {\it St\"uckelberg - Petermann RG ${\cal R}$} \cite{SP} is the version of the RG
as it appears in causal perturbation theory \cite{PStora,HW3,DFret,DFbros,qap,BDF}. It relies on the
non-uniqueness of the $S$-matrix ${\bf S}$. Roughly speaking the Main Theorem 
states that a change ${\bf S}\rightarrow {\bf \hat S}$ of the renormalization presription
can be absorbed in a renormalization of the interaction $V\rightarrow Z(V)$ \cite{PStora}:
\begin{equation}
 {\bf \hat S}(V)={\bf S}(Z(V))\quad \forall V\in {\cal F}_{\rm loc} \label{def:Z}
\end{equation}
(where ${\cal F}_{\rm loc}$ is the space of all local interactions).
The St\"uckelberg - Petermann RG ${\cal R}$ is the set of all bijective maps $Z:{\cal F}_{\rm loc}\to
{\cal F}_{\rm loc}$ appearing in this relation when ${\bf S}$ and ${\bf \hat S}$ run through all
admissible $S$-matrices. From the complete statement of the Main Theorem \cite{DFret,DFbros} (see Sect.~3.2)
it follows that ${\cal R}$ is indeed a group; due to \eqref{def:Z} 
${\cal R}$ can be interpreted as the group of finite 
renormalizations of the $S$-matrix.
\item The {\it RG in the sense of Wilson} relies usually on a functional integral approach, it describes the
dependence of the theory on a cutoff $\Lambda$, which one introduces to avoid UV-divergences.
We imitate this approach in the framework of causal perturbation theory by proceeding as follows \cite{BDF}:
Let $p_\Lambda$ be a regularized Feynman propagator, which converges (in an appropriate sense) 
to the Feynman propagator for $\Lambda\to\infty$. In terms of $p_\Lambda$ we construct 
a regularized $S$-matrix ${\bf S}_\Lambda$. With that we define
the {\it effective potential $V_\Lambda$ at scale $\Lambda$} as a function
of the original interaction $V$ by the condition that the cutoff theory with interaction $V_\Lambda$
agrees with the exact theory with interaction V, that is 
\beq
{\bf S}_\Lambda(V_\Lambda)={\bf S}(V)\qqbox{or explicitly} 
V_\Lambda := {\bf S}_\Lambda^{-1}\circ {\bf S}\,(V)\ .\label{def:effpot}
\eeq
Here we use that ${\bf S}_\Lambda$ is invertible (see \cite{BDF} and Sect.~4).
We point out that in general $V_\Lambda$ is a {\it non-local} interaction.

The problem with this definition is that usually ${\bf S}$ is unknown. Therefore, one computes $V_\Lambda$ 
by solving {\it Polchinski's flow equation} \cite{Pol,Salm,KKS}, 
which can be derived from the definition
\eqref{def:effpot} (see Sect.~5 and \cite{BDF}) and has the form
\begin{equation}
\frac{d}{d\Lambda}V_\Lambda= F_\Lambda(V_\Lambda\otimes V_\Lambda)\ ,
\end{equation}
where $F_\Lambda$ is linear and explicitly known.
Integration of the flow equation yields $V_\Lambda$ and from that the $S$-matrix is obtained by
${\bf S}(V)={\bf S}_\Lambda(V_\Lambda)$; in practice it is easier to compute
\begin{equation}
\lim_{\Lambda\to\infty}{\bf S}_\Lambda(V_\Lambda)\ ,
\end{equation}
since several terms vanish in this limit and, hence, need not to be computed in detail.

We point out that there is no group structure in the mathematical sense in Wilson's RG. 
\end{itemize}

\section{Star-product quantization}\setcounter{equation}{0}

We use a formalism which arises 
when applying deformation quantization to the underlying free classical fields \cite{DFloop,DFsiena}.

To simplify the notations we restrict this paper to a real scalar field $\varphi$ in $d$-dimensional
Minkowski space. We work with an ``off-shell formalism'' which means that the classical
field configuration space is $C^\infty(\RR^d)$ (and not only the space of solutions of the field equations).

We define the space ${\cal F}$ of
{\bf observables} as the set of all functionals $F: \>C^\infty(\RR^d)\rightarrow \CC$, 
which are infinitely differentiable and all functional derivatives $\frac{\delta^n F}{\delta\varphi^n}$ 
($n\in\NN$) must be distributions with compact supports. There is an 
additional defining condition on the wave front sets ${\rm WF}(\frac{\delta^n F}{\delta\varphi^n})$, 
$n\in\NN$, which is a 
microlocal version of translation invariance \cite{DFloop,DFsiena,DFret,BDF}.

The following subspaces of ${\cal F}$ will be of crucial importance.
\begin{itemize}
\item The {\it non-local functionals} ${\cal F}_0$ are defined by the stronger requirement
that $\frac{\delta^n F}{\delta\varphi^n}$ is a smooth function with compact support for all $n\in\NN$.
\item The {\it local functionals} ${\cal F}_{\rm loc}$ are defined by the
additional condition that
$\frac{\delta^n F}{\delta\varphi^n}(x_1,\dots, x_n) =0$ if $x_i\neq x_j$ for some $(i,j)$.
\end{itemize}

{\it Local} interactions are usually of the form 
\beq
F(\varphi)=\int dx\,f(x)\,L(\varphi(x),\partial\varphi(x),\partial\partial\varphi(x),...)\in {\cal F}_{\rm loc}\label{interaction}
\eeq
where $f\in {\cal D}(\RR^d)$ switches the interaction
and $L\in {\cal C}^\infty$ does not need to be a polynomial. 

By the support of an observable $F$ we mean
${\rm supp}\>F:= {\rm supp}\>\frac{\delta F}{\delta\varphi}\ $.

The {\it pointwise product}
\beq
(F\cdot G)(\varphi):= F(\varphi)\cdot G(\varphi)
\eeq
is commutative, we call it also the {\it classical} product.

To obtain the {\bf Poisson algebra of free fields} we define the Poisson bracket.
For this purpose we need the retarded propagator $\Delta_R$ of the Klein-Gordon operator, 
from which we construct the commutator function 
\beq
\Delta(x):=\Delta_R(x)-\Delta_R(-x)=-\Delta(-x)\ .
\eeq 
Graphically the Poisson bracket of $F,G\in {\cal F}$ is defined by contracting once $F$ with $G$
with propagator $\Delta$, i.e.
\begin{equation}
\{F,G\}:=\int dx\,dy\,\frac{\delta F}{\delta\varphi(x)}\,\Delta(x-y)\,\frac{\delta G}{\delta\varphi(y)}\ .
\label{PB}
\end{equation} 

In view of a $\star$-product quantization we introduce ${\cal F}[[\hbar]]$ as 
the space of formal power series in $\hbar$ with coefficients in ${\cal F}$
(and similar for subspaces of ${\cal F}$).  

Now we define the {\bf 'product with propagator $p$'}
\beq
\star_p:{\cal F}_1[[\hbar]]\times{\cal F}_1[[\hbar]]\to {\cal F}[[\hbar]]\,;\;(F,G)\mapsto F\star_p G\ ,
\eeq
(where ${\cal F}_1$ is a subspace of ${\cal F}$ and
$p\in {\cal S}'(\RR^d)$ is a distribution with suitable properties, see the examples below)
by the prescription
\begin{align}
 F\star_p G:=&\sum_{n\geq 0}\frac{\hbar^n}{n!}\int dx_1...dy_1...
\frac{\delta^n F}{\delta\varphi(x_1)...\delta\varphi(x_n)}\,\notag\\
& p(x_1-y_1)...p(x_n-y_n)\,\frac{\delta^n G}{\delta\varphi(y_1)...\delta\varphi(y_n)}\ .\label{product-p}
\end{align}
The zeroth term ($n=0$) is the classical product.
The $n$-th term has prercisely $n$ contractions of $F$ with $G$, each contraction with the 
propagator $p$. Hence, $F\star_p G$ is the sum over all possible contractions of $F$ with $G$.
For non-local functionals (${\cal F}_1= {\cal F}_0$) the integral in \eqref{product-p}
exists for all $p\in {\cal S}'(\RR^d)$
(it means smearing of the distribution $p\otimes...\otimes p$ with the test function 
$\frac{\delta^n F}{\delta\varphi(x_1)...}\cdot\frac{\delta^n G}{\delta\varphi(y_1)...}$).
One can show that $\star_p$ is associative, obviously it is distributive.

The following examples for $\star_p$ are of crucial importance.
\begin{itemize}
\item {\bf $\star$-product quantization.} We choose for $p$ a Hadamard function $H\equiv H_m$.
A Hadamard function is a Poincar\'e invariant solution of the Klein Gordon equation with mass $m$,
which is smooth in $m\geq 0$ and fulfils
\begin{itemize} 
\item[(i)] $H(z)-H(-z)=i\Delta(z)\quad$ and
\item[(ii)] $z\mapsto (H(z)-\Delta_+(z))\quad$ is smooth, where $\Delta_+$ is the Wightman $2$-point function.
Since powers $(\Delta_+(z))^n$ exist, this condition implies that $(H(z))^n $ exists $\forall n\in\NN$.
\end{itemize}
$F\star_H G$ exists also for local functionals,
i.e.~it exists $\forall F,G\in {\cal F}$. Namely, property (ii) and the wave 
front set property of $\frac{\delta^n F}{\delta\varphi^n}$ and $\frac{\delta^n G}{\delta\varphi^n}$
imply that the appearing products of distributions exist \cite{DFloop}.

The product $\star_H$ viewed as a map ${\cal F}\times{\cal F}\to {\cal F}[[\hbar]]$ 
(i.e.~the arguments are $\sim\hbar^0$)
is a $\star$-product, i.e. $F\star_H G$ is a $\hbar$-dependent deformation of $F\cdot G$,
\begin{equation}
\lim_{\hbar\to 0}F\star_H G=F\cdot G\ ,
\end{equation}
with
\begin{equation}
\lim_{\hbar\to 0}\frac 1{i\hbar}(F\star_H G-G\star_H F)=\{F,G\}\ .\label{star-prod-2}
\end{equation}
The validity of the last relation relies on property (i) of $H$. 
The Wightman $2$-point function $\Delta_+$ yields also a $\star$-product 
$\star_{\Delta_+}$, but $\Delta_+$ is not smooth in $m$ at $m=0$.

\item{\bf Time ordered product of nonlocal functionals.} 
The time ordered product with respect to the $\star$-product $\star_H$ must satisfy
\beq
T(\varphi(x)\varphi(y)):=
\begin{cases}\varphi(x)\star_H\varphi(y) &{\rm if}\quad x^0>y^0\\
\varphi(y)\star_H\varphi(x) &{\rm if}\quad y^0>x^0\end{cases}
\eeq
These two cases can be summarized as
\beq
T(\varphi(x)\varphi(y))=\varphi(x)\star_{H_F}\varphi(y)
\eeq
with 
\beq
H_F(z):= \Theta (z^0)H(z)+\Theta (-z^0)H(-z)(=H_F(-z))\ .
\eeq
Hence, the propagator for the time ordered product w.r.t. $\star_H$ is $H_F$ and, therefore, we define
the time ordered product of {\bf non-local} functionals as the product with propagator $H_F$:
\beq
 T(F_1\otimes ...\otimes F_n):=F_1\star_{H_F}...\star_{H_F}F_n\ ,\quad \forall \, F_1,...,F_n\in {\cal F}_0\ .
\label{def:Tprod}
\eeq 

Since $H_F(-z)=H_F(z)$, this time ordered product is commutative, hence it cannot be a 
$\star$-product (due to \eqref{star-prod-2}).
\end{itemize}

\section{The St\"uckelberg - Petermann renormalization group} \setcounter{equation}{0}
 
\subsection{Time ordered product of local functionals} 
For simplicity we assume from now on that all observables $F\in {\cal F}$ 
are polynomial in $\varphi$ and its derivatives, and we write ${\cal F}$ for ${\cal F}[[\hbar]]$ (and similar 
for subspaces of ${\cal F}$).

Trying to extend the definition \eqref{def:Tprod} to local functionals, powers
$(H_F(z))^n$ and terms like $H_F(z_1)H_F(z_2)H_F(z_1+z_2)$ appear, which do not exist
in $d\geq 4$ or $d\geq 6$ dimensions! (These 
are the famous UV-divergences of perturbative QFT.)
Therefore, we define the time ordered product of local functionals in an alternative, axiomatic way:
we require that the time ordered product of $n$-th order
\beq 
T_n:{\cal F}_{\rm loc}^{\otimes n}\rightarrow {\cal F} 
\eeq
is a linear and totally symmetric map. With that the defining axioms can be given in terms of the 
generating functional -- the $S$-matrix 
\beq 
{\bf S}:{\cal F}_{\rm loc}\rightarrow {\cal F}\, ;\,{\bf S}(V):=\sum_{n=0}^\infty\frac{T_n(V^{\otimes n})}{n!}\ . 
\eeq 
Or vice versa $T_n$ is obtained from ${\bf S}$ by
\beq
T_n(V^{\otimes n})={\bf S}^{(n)}(0)(V^{\otimes n})\equiv\frac{d^n}{d\lambda^n}{\bf S}(\lambda V)|_{\lambda=0}
\eeq
(where ${\bf S}^{(n)}(0)$ denotes the $n$-th derivative of ${\bf S}$ at the origin).
We use the axioms of causal perturbation theory \cite{EG,Sto,BF}\footnote{In view of the generalization to curved 
spacetimes we work with a somewhat modified version of the axioms given in \cite{DFret}.}
\begin{description}
\item [Causality:] ${\bf S}(A+B)={\bf S}(A)\star_{H_m} {\bf S}(B)$ if ${\rm supp}\>A$ is later than  ${\rm supp}\>B$,
i.e. $\supp A\cap (\supp B+\bar V_-)=\emptyset$ (where $\bar V_-$ denotes the full and closed backward light cone). 
\item[Starting element:] ${\bf S}(0)=1$, ${\bf S}^{(1)}(0)={\rm id}\  $.
\item[Field Independence:]  $\delta {\bf S}/\delta\varphi=0\ $.
\item [Poincar\'e invariance] 
\item [Unitarity:] $\overline{{\bf S}(-V)}\star_{H_m} {\bf S}(\overline{V})=1\ $ (where the bar means complex conjugation).
\item [Smoothness in $m$:] ${\bf S}$ depends smoothly on the mass $m$ of the free theory $\forall m\geq 0$.
\item [Scaling:] ${\bf S}$ scales almost homogeneously under $(x,m)\mapsto (\rho x,\rho^{-1}m)$, by which we mean
that homogeneous scaling (holding for the corresponding classical theory)
is maintained up to powers of ${\rm log}\>\rho$.  
\end{description}
Note that in the Causality condition the $\star$-product w.r.t.~a Hadamard function $H_m$ appears. It is here 
where the information about the free field equation (in particular about the value of the mass $m\geq 0$) enters
the axioms. If $H$ would be replaced by $\Delta_+$, Smoothness in $m$ would be violated at $m=0$.

Epstein and Glaser showed that these axioms have a solution \cite{EG} (for somewhat alternative
proofs see \cite{Sto,BF,DFret}):
they gave a construction of the time ordered products $T_n$ by induction on $n$. 
In this construction renormalization appears as the problem of extending the distributional kernels of $T_n$  
from ${\cal D}(\RR^{dn}\setminus\Delta_n)$ to ${\cal D}(\RR^{dn})$ 
where $\Delta_n:=\{(x_1,...,x_n)\in\RR^{dn}\,|
\,x_1=x_2=...=x_n\}$. The non-uniqueness of this extension is 
the reason for the non-uniqueness of the $S$-matrix.

'Causality' and 'Starting element' are the basic axioms; the other axioms are 
not mandatory, they are called
(re)normalization conditions because their only purpose is to restrict the set of admissible extensions.

\subsection{Non-uniqueness of the $S$-matrix}

We define the {\bf St\"uckelberg - Petermann RG ${\cal R}$} as the set of all analytic bijections  
$Z\>:\>{\cal F}_{\rm loc}\rightarrow {\cal F}_{\rm loc}$ with 
\begin{description}
\item [Starting element:] $Z(0)=0\,,\quad Z^{(1)}(0)  = {\rm id}\,,\quad  Z = {\rm id} + O(\hbar)\ $.
\item [Locality:] $Z$ is local in the sense that\\ 
$Z(A+B+C)=Z(A+B)-Z(B)+Z(B+C)$\\
 if ${\rm supp}\> A\cap{\rm supp}\> C=\emptyset\ $.
\item [Field Independence:] $\delta Z/\delta\varphi=0\ $.
\item [Poincar\'e invariance]
\item [Unitarity:] $\ovl{Z}(-V)+Z(V)=0\ $.
\item [Smoothness in $m$:] $Z$ depends smoothly on $m\geq 0\ $.
\item [Scaling:] $Z$ scales almost homogeneously under $(x,m)\mapsto (\rho x,\rho^{-1}m)\ $.
\end{description}
Every renormalization condition on ${\bf S}$ has a corresponding requirement on $Z$. Analyticity of $Z$ means 
that it is given by its Taylor series:
\beq
Z(V)=\sum_{n=1}^\infty\frac{Z^{(n)}(0)(V^{\otimes n})}{n!}\ .
\eeq

The 'Main Theorem' describes the non-uniqueness of the $S$-matrix in terms of the 
St\"uckelberg - Petermann RG ${\cal R}$.
\begin{thm} (i) Given two renormalization prescriptions ${\bf S}$ and ${\bf \hat S}$
there exists a unique map $Z:{\cal F}_{\rm loc}\to {\cal F}_{\rm loc}$ with $Z(0)=0$ and
\beq
{\bf \hat S}={\bf S}\circ Z\ .
\eeq
This $Z$ is an element of the St\"uckelberg - Petermann RG ${\cal R}$.\\ 
(ii) Conversely, given an $S$-matrix ${\bf S}$ and an arbitrary $Z\in {\cal R}$, then
${\bf \hat S}:= {\bf S}\circ Z$ satisfies also the axioms for an $S$-matrix.
\end{thm}
For the proof we refer to \cite{DFret}.

A Corollary of this Theorem states that, for $Z_1,Z_2\in {\cal R}$, 
the composition $Z_1\circ Z_2$ is also an element of ${\cal R}$,
i.e.~that ${\cal R}$ is indeed a {\it group} \cite{DFret}. Namely, given $Z_1,Z_2\in {\cal R}$ and 
choosing an arbitrary $S$-matrix ${\bf S}$,
part (ii) implies that ${\bf S}_1:= {\bf S}\circ Z_1$ and ${\bf S}_2:= {\bf S}_1\circ Z_2$ satisfy also the axioms.
From ${\bf S}_2= {\bf S}\circ(Z_1\circ Z_2)$ and part (i) it follows that $Z_1\circ Z_2\in {\cal R}$.

\section{Regularized time-ordered product}\setcounter{equation}{0}
The definition \eqref{def:Tprod} of the time ordered product of non-local functionals 
can be extended to local functionals, if one regularizes the Feynman propagator $H_F$ by
introducing a cutoff $\Lambda$. 

Let $(p_\Lambda)_{\Lambda>0}$ be a family of {\it test functions} 
($p_\Lambda\in{\cal S}(\RR^d)$) which approximates $H_F$, more precisely
$$
\lim_{\Lambda\to\infty}p_\Lambda= H_F\qbox{in the H\"ormander topology \cite{Hoerm},}
$$
and for $\Lambda\to 0$ it is required that
$$
p_0=0 \qbox{or} \lim_{\Lambda\to 0}p_\Lambda= 0\qbox{in the H\"ormander topology.}
$$
With regard to $H_F(-z)=H_F(z)$, we additionally require $p_\Lambda(-z)=p_\Lambda(z)$.

The regularized time-ordered product,
\begin{equation}\label{T-reg}
T_\Lambda(F^{\otimes n}):= F\star_{p_\Lambda}...\star_{p_\Lambda} F
\end{equation}
is well-defined $\forall F\in {\cal F}$ since $p_\Lambda\in {\cal S}(\RR^d)\ $.

The corresponding generating functional  is the regularized $S$-matrix 
\begin{equation}\label{S-reg}
{\bf S}_\Lambda\>:\>{\cal F}\rightarrow {\cal F}\,;\,{\bf S}_\Lambda(F):=\sum_{n=0}^\infty\frac{1}{n!}\,T_\Lambda(F^{\otimes n})
=: e_{\star_{p_\Lambda}}^F
\end{equation}
(The last expression is a suggestive short-hand notation for the series.) In contrast to the exact $S$-matrix 
${\bf S}$, the domain of ${\bf S}_\Lambda$ is ${\cal F}$ (and not only ${\cal F}_{\rm loc}$) and 
 ${\bf S}_\Lambda$ is invertible. We also point out that $\lim_{\Lambda\to\infty}\,{\bf S}_\Lambda$
does not exist in general.\\
{\bf Proof of invertability:}
following \cite{BDF} we write the product $\star_{p_\Lambda}$ alternatively as
\beq\label{product-alternative}
F\star_{p_\Lambda} G=\tau_\Lambda\,(\tau_\Lambda^{-1}F\cdot\tau_\Lambda^{-1}G)\ ,
\eeq
where
\begin{equation}\label{tau}
\tau_\Lambda F:=\exp({i\hbar \Gamma_\Lambda})\, F
\end{equation}
with 
\begin{equation}\label{Gamma}
\Gamma_{\Lambda}:=\frac12\int dx\,dy\,\, p_\Lambda(x-y)\,\frac{\delta^2}{\delta\varphi(x)\delta\varphi(y)} \ .
\end{equation}
Graphically $(\tau_\Lambda F)(\varphi)$ is the sum over all possible contractions 
(with propagator $p_\Lambda$) of $\varphi$ in 
$F(\varphi)\ $. Obviously, the inverse operator $\tau_\Lambda^{-1}$ is $\tau_\Lambda^{-1}=\exp({-i\hbar \Gamma_\Lambda})\ $.
Note that the operators $\tau_\Lambda^{-1}$ in \eqref{product-alternative} are needed to remove
the tadepole diagrams.

With \eqref{product-alternative} ${\bf S}_\Lambda$ can be written as
\beq\label{S-tau}
{\bf S}_\Lambda =\tau_\Lambda\circ\mathrm{exp}\circ\tau_\Lambda^{-1}
\eeq
(where $\mathrm{exp}\, F=1+F+F\cdot F/2!+...$)
from which  it is obvious that ${\bf S}_\Lambda$ is invertible:
\beq
{\bf S}_\Lambda^{-1}=\tau_\Lambda\circ\mathrm{log}\circ\tau_\Lambda^{-1}\ .\quad\square\label{SLa-1}
\eeq
{\bf Examples} for regularized (Feynman) propagators:
\begin{itemize}
\item {\bf Euklidean theory with mass $m>0$} (following \cite{Salm}). \\
Let $K\in {\cal C}^\infty(\RR_0^+,[0,1])$
with 
$$K(x)=\begin{cases}0 &{\rm if}\quad x\geq 4\\
1 &{\rm if}\quad x\leq 1\end{cases}
$$
and $K'(x)<0\,\,\forall x\in(1,4)$, i.e. $K$ is a smooth version of a step function. The Eulklidean
propagator is regularized by cutting off the momenta above a scale $\Lambda$:
\beq\label{prop-eukl}
\hat p_\Lambda(k):=\frac 1{(2\pi)^2\,(k^2+m^2)}\,K(\frac{k^2}{\Lambda^2})\in {\cal S}(\RR^d)\quad 
(k^2\equiv k_0^2+\vec k^2)\ .
\eeq
It follows $p_\Lambda(x)\in {\cal S}(\RR^d)$, $p_\Lambda(-x)=p_\Lambda(x)$
and that 
\beq
\lim_{\Lambda\to\infty}\hat p_\Lambda(k)=\frac 1{(2\pi)^2\,(k^2+m^2)}\qbox{(=Eukl. prop.),}
\lim_{\Lambda\to 0}\hat p_\Lambda(k)= 0
\eeq 
w.r.t.~the weak topology of ${\cal S}'(\RR^d)$. Due to the continuity of the inverse Fourier transformation
from ${\cal S}'(\RR^d)$ to ${\cal S}'(\RR^d)$, these convergence statements hold also in $x$-space w.r.t.~the weak topology.
They are valid also w.r.t.~the H\"ormander topology in $x$-space. For the Euklidean propagator $p(x):=\lim_{\Lambda\to\infty}p_\Lambda(x)$
one obtains
$$
p(x)=\frac{2\,\pi^{\frac{d+1}{2}}}{\Gamma(\frac{d-1}{2})}\,
\int_m^\infty dq\,(q^2-m^2)^{\frac{d-3}{2}}\,e^{-q\,|x|}\in {\cal S}'(\RR^d)\ ,
$$
where $|x|\equiv \sqrt{x_0^2+\vec x^2}$.
In low dimensions the remaining integral gives
$$
d=2\, :\quad p(x)=2\pi\, K_0(m\,|x|)
$$
(where $K_0$ is a modified Bessel function of second kind) and
$$ 
d=3\, :\quad p(x)=2\,\pi^2\,\frac{e^{-m\,|x|}}{|x|}\ . 
$$ 
Both expressions have an integrable singularity at $x=0$ and decay exponentially for $|x|\to\infty$.
\item {\bf $\epsilon$-regularized relativistic theory with $m>0$} (following \cite{KKS}).\\
$\epsilon$-regularization of the relativistic theory means that the Minkowski metric is replaced by $i\eta_\epsilon$,
where
$$
k\eta_\epsilon k:=k_0^2\,(\epsilon-i)+\vec k^2\,(\epsilon+i)\qbox{with} \epsilon >0\ .
$$
Note that $k\eta_\epsilon k\not= 0$ for $k\not= 0$ and that
\beq
\mathrm{Re}(k\eta_\epsilon k+(\epsilon+i)m^2)=\epsilon\,(k_0^2+\vec k^2+m^2)
\geq\epsilon\, m^2\quad\forall k\ .\label{eMink-prop}
\eeq
For the Feynman propagator of the $\epsilon$-regularized relativistic theory,
$$
\hat p_\epsilon(k)=\frac{i}{(2\pi)^2\,(k\eta_\epsilon k+(\epsilon+i)m^2)}
$$
an UV-cutoff $\Lambda$ is introduced by an exponential damping:
\beq
\hat p_{\epsilon,\Lambda}(k):=e^{-\Lambda^{-1}(k\eta_\epsilon k+(\epsilon+i)m^2)}\,\hat p_\epsilon(k)=\frac{i}{(2\pi)^2}\,
\int_{\Lambda^{-1}}^\infty d\alpha\,e^{-\alpha (k\eta_\epsilon k+(\epsilon+i)m^2)}\ .
\eeq
Obviously it holds $\hat p_{\epsilon,\Lambda}(k)\in {\cal S}(\RR^d)$ and hence $p_{\epsilon,\Lambda}(x)\in {\cal S}(\RR^d)$.
We also see that $p_{\epsilon,\Lambda}(-x)=p_{\epsilon,\Lambda}(x)$. 

For fixed $\epsilon >0$ we find
$$ 
\lim_{\Lambda\to\infty} p_{\epsilon,\Lambda}(x)=p_{\epsilon}(x)\qqbox{and}  
\lim_{\Lambda\to 0} p_{\epsilon,\Lambda}(x)=0  
$$
w.r.t.~the weak topology of ${\cal S}'(\RR^d)$ and also w.r.t.~the H\"ormander topology. Namely, for $\Lambda\to\infty$
the convergence behaviour is essentially similar to the Euklidean case treated above (due to \eqref{eMink-prop})
and for $\Lambda\to 0$ the behaviour of $p_{\epsilon,\Lambda}(x)$ is dominated by a prefactor
$e^{-\Lambda^{-1}\,\epsilon\,m^2}$.

For $\epsilon\downarrow 0$ the family $(\hat p_\epsilon(k))_{\epsilon >0}$ of analytic functions
converges to the distribution
$$
\lim_{\epsilon\downarrow 0}\hat p_{\epsilon}(k)=\frac{1}{(2\pi)^2\,(m^2-k^2-i0)}\qbox{(=Feynman propagator),} 
$$ 
w.r.t.~the weak topology of ${\cal S}'(\RR^d)$ and, hence, this holds also in $x$-space:
$\lim_{\epsilon\downarrow 0} p_{\epsilon}(x)=$(Feynman propagator) in ${\cal S}'(\RR^d)$.
Whether this holds also w.r.t.~the H\"ormander topology is a more difficult question, which cannot be answered with the 
mathematical tools explained in this paper.
\end{itemize}

In both examples the propagator
\beq\label{prop-UV-IR}
p_{\Lambda,\Lambda_0}:=p_{\Lambda_0}-p_{\Lambda} \quad (0<\Lambda\leq\Lambda_0<\infty)
\eeq
has an UV-cutoff ($\Lambda_0\to\infty$) and an IR-cutoff ($\Lambda\to 0$).

\section{Effective potential and flow equation}\setcounter{equation}{0}
To define the effective potential $V_\Lambda$ we recall that ${\bf S}_\Lambda$ is explicitly known and invertible and that 
${\bf S}$ exits (although it is usually unknown). With that the effective potential $V_\Lambda$ at scale $\Lambda$
can be defined as explained in the introduction:
\begin{equation}\label{eff-pot}
V_\Lambda:= {\bf S}_\Lambda^{-1}\circ {\bf S}\, (V)\ .
\end{equation}
We also recall that in general $V_\Lambda\not\in {\cal F}_{\rm loc}$.

Similarly to ${\bf S}(V)$, $V_\Lambda$ can be viewed as a formal power series in $\hbar$, or in $V$, or in both.
For the lowest terms of the expansion in $V$ we obtain
\beq
V_\Lambda =V+{\cal O}(V^2)\ ,\label{V-expansion}
\eeq
by using the axiom Starting element and \eqref{SLa-1}.

In particular for $\Lambda=0$ we obtain
\beq 
V_0=\mathrm{log}\circ {\bf S}(V)\label{La=0}
\eeq 
(due to ${\bf S}_0(V)=e^V\equiv\sum_n\frac{V\cdot ...\cdot V}{n!}$). 

{\bf Flow operator.} From the definition \eqref{eff-pot} it follows
\beq
V_\Lambda={\bf S}_\Lambda^{-1}\circ {\bf S}_{\Lambda_0}\, (V_{\Lambda_0})\ ,\label{flow}
\eeq
i.e. ${\bf S}_\Lambda^{-1}\circ {\bf S}_{\Lambda_0}$ is 
the ``flow of the effective potential from $\Lambda_0$ to $\Lambda$''.

We want to clarify the relation between this flow operator and ${\bf S}_{\Lambda,\Lambda_0}$;
by the latter we mean the regularized $S$-matrix with propagator $p_{\Lambda,\Lambda_0}\ $ \eqref{prop-UV-IR}, 
analogously to \eqref{T-reg} and \eqref{S-reg}. 
Similarly to \eqref{S-tau}, ${\bf S}_{\Lambda,\Lambda_0}$ satisfies
\beq
{\bf S}_{\Lambda,\Lambda_0}=\tau_{\Lambda,\Lambda_0}\circ
\mathrm{exp}\circ\tau_{\Lambda,\Lambda_0}^{-1}\ ,\label{S-1-tau}
\eeq
where $\tau_{\Lambda,\Lambda_0}$ is defined as $\tau_\Lambda$ \eqref{tau}, with propagator
$p_{\Lambda,\Lambda_0}$ instead of $p_\Lambda$.
For $\Lambda=0$ we have $p_0=0$, $p_{0,\Lambda_0}= p_{\Lambda_0}$,
hence $\tau_0=\mathrm{id}$, $S_0^{-1}=\log$ and $S_{0,\Lambda_0}=S_{\Lambda_0}$.
With that, ${\bf S}_0^{-1}\circ {\bf  S}_{\Lambda_0}$ (i.e.~the flow from $\Lambda_0$ to $0$) is equal to  
$\log\circ {\bf S}_{0,\Lambda_0}\ $. For the flow from $\Lambda_0$ to an arbitrary $\Lambda\in [0,\Lambda_0]$
we assert
\beq\label{rel-flow-op}
{\bf S}_\Lambda^{-1}\circ {\bf S}_{\Lambda_0}=
\tau_\Lambda\circ\mathrm{log}\circ {\bf S}_{\Lambda,\Lambda_0}\circ\tau_\Lambda^{-1}\ .
\eeq
{\bf Proof:} We first note that
$$
\tau_{\Lambda,\Lambda_0}=\exp(i\hbar (\Gamma_{\Lambda_0}-\Gamma_\Lambda))=
\tau_\Lambda^{-1}\circ\tau_{\Lambda_0}=\tau_{\Lambda_0}\circ\tau_\Lambda^{-1}\ .
$$
Due to \eqref{S-tau} and \eqref{S-1-tau} it holds
\beq\label{SLL0}
{\bf S}_{\Lambda,\Lambda_0}=\tau_{\Lambda,\Lambda_0}\circ\mathrm{exp}\circ\tau_{\Lambda,\Lambda_0}^{-1}
=\tau_\Lambda^{-1}\circ {\bf S}_{\Lambda_0}\circ \tau_\Lambda
\eeq 
and with that we obtain
$$
{\bf S}_\Lambda^{-1}\circ {\bf S}_{\Lambda_0}\circ\tau_\Lambda =\tau_\Lambda\circ\mathrm{log}\circ\tau_\Lambda^{-1}
\circ {\bf S}_{\Lambda_0}\circ \tau_\Lambda=\tau_\Lambda\circ\mathrm{log}\circ {\bf S}_{\Lambda,\Lambda_0}\ .\quad\square
$$

{\bf Flow equation.} The flow equation (cf.~\cite{Pol,Salm,KKS,BDF}) is a differential equation for $V_\Lambda$
as a function of $\Lambda$.
\begin{thm}
\begin{eqnarray}
 \frac{d}{d\Lambda}V_\Lambda=& -\frac{\hbar}2 \int dx\,dy\,\frac{d\,p_\Lambda(x-y)}{d\Lambda}\,
\frac{\delta V_\Lambda}{\delta\varphi(x)}\star_{p_\Lambda}\frac{\delta V_\Lambda}{\delta\varphi(y)}\label{floweq1}\\
=& -\frac 12 \frac{d}{d\lambda}\vert_{\lambda=\Lambda}(V_\Lambda\star_{p_{\lambda}}V_\Lambda)\label{floweq}
\end{eqnarray}
\end{thm}

{\bf Proof}\footnote{A somewhat different proof is given in \cite{BDF}.}. From ${\bf S}_\Lambda(V)= e_{\star_{p_\Lambda}}^V$
we see that
$$
\frac{d}{d\lambda}{\bf S}_\Lambda(V_\lambda)= \frac{d\,V_\lambda}{d\lambda}\star_{p_\Lambda}{\bf S}_\Lambda(V_\lambda)
$$
and with that we obtain
\begin{equation*}
0=\frac{d}{d\Lambda}{\bf S}_\Lambda(V_\Lambda)=\frac{d}{d\lambda}\vert_{\lambda=\Lambda}{\bf S}_\lambda(V_\Lambda)
+ \frac{d\,V_\Lambda}{d\Lambda}\star_{p_\Lambda}{\bf S}_\Lambda(V_\Lambda)
\end{equation*}
Due to ${\bf S}_\Lambda(F)=1+{\cal O}(F)$ the inverse (w.r.t.~$\star_{p_\Lambda}$) regularized $S$-matrix ${\bf S}_\Lambda(F)^{-1}$ 
exists. With that it follows that
\beq
\frac{d\,V_\Lambda}{d\Lambda}=-\frac{d}{d\lambda}\vert_{\lambda=\Lambda}{\bf S}_\lambda(V_\Lambda)
\star_{p_\Lambda}{\bf S}_\Lambda(V_\Lambda)^{-1}\ .\label{flussgl-beweis}
\eeq
From the definition \eqref{product-p} of $\star_p$ we obtain
\begin{equation}
\frac{d}{d\Lambda}\frac{F\star_{p_\Lambda}F}2 =
\frac{\hbar}2 \int dx\,dy\,\frac{d\,p_\Lambda(x-y)}{d\Lambda}\,
\frac{\delta F}{\delta\varphi(x)}\star_{p_\Lambda}\frac{\delta F}{\delta\varphi(y)}\label{D*prod}
\end{equation}
and for $n$ factors
\begin{equation*}
\frac{d}{d\Lambda}\frac{T_\Lambda(F^{\otimes n})}{n!} =
\frac{\hbar}{2\,(n-2)!} \int dx\,dy\,\frac{d\,p_\Lambda(x-y)}{d\Lambda}\,
\frac{\delta F}{\delta\varphi(x)}\star_{p_\Lambda}\frac{\delta F}{\delta\varphi(y)}
\star_{p_\Lambda}F\star_{p_\Lambda}...\star_{p_\Lambda}F
\end{equation*}
(2 factors $\frac{\delta F}{\delta\varphi}$ and $(n-2)$ factors $F$). Summing over $n$ we obtain
\beq
\frac{d}{d\Lambda}{\bf S}_\Lambda(F)=\frac{\hbar}2 \int dx\,dy\,\frac{d\,p_\Lambda(x-y)}{d\Lambda}\,
\frac{\delta F}{\delta\varphi(x)}\star_{p_\Lambda}\frac{\delta F}{\delta\varphi(y)}
\star_{p_\Lambda}{\bf S}_\Lambda(F)\ .
\eeq
Inserting this into \eqref{flussgl-beweis} it results \eqref{floweq1}, from which we obtain \eqref{floweq}
by using \eqref{D*prod}. $\quad\square$


{\bf Construction of $V_\Lambda$.} Usually ${\bf S}$ is unknown, only $V$ and $(p_\Lambda)_{\Lambda>0}$ are given,
and from that $V_\Lambda$ is computed by solving the flow equation. In perturbation theory this amounts to an inductive
construction of $V_\Lambda$ as a formal power series in $V$. Namely, denoting by $V_\Lambda^{(n)}$ the 
term in $V_\Lambda$ of order $n$ in $V$, and taking $V_\Lambda^{(0)}=0$ into account, the 
perturbative version of the flow equation reads 
\begin{equation} 
 \frac{d}{d\Lambda}V_\Lambda^{(n)}= \sum_{k=1}^{n-1} 
 -\frac 12 \frac{d}{d\lambda}\vert_{\lambda=\Lambda}(V_\Lambda^{(k)}\star_{p_{\lambda}} 
V_\Lambda^{(n-k)})\ . 
\end{equation} 
Proceeding inductively, we start with $V_\Lambda^{(1)}=V$ \eqref{V-expansion} and
assume that $V_\Lambda^{(k)}$ is known for all $k<n$. Then, the r.h.s.~is known and, hence, an integration yields 
$V_\Lambda^{(n)}$. A major problem is the determination of the integration constant by a suitable boundary value.  
(The value \eqref{La=0} at $\Lambda=0$ does not help, because it contains the unknown ${\bf S}$.) 
We refer to the usual procedure which is roughly sketched in the next section.

Concerning the removal of the cutoff $\Lambda$, we point out that $V_\Lambda$ diverges in 
general for $\Lambda\to\infty$. But $\lim_{\Lambda\to\infty}{\bf S}_\Lambda(V_\Lambda)$
exists and gives ${\bf S}(V)$. 

\section{Comparison with the functional integral approach}\setcounter{equation}{0}
First we roughly sketch the usual procedure for the Euklidean theory, following \cite{Salm}.
One defines an effective action $G_{\Lambda,\Lambda_0}$ by the functional integral
\beq
e^{G_{\Lambda,\Lambda_0}(\varphi)}:=\int d\mu_{p_{\Lambda,\Lambda_0}}(\phi)\,
e^{-\lambda V^{(\Lambda_0)}(\phi+\varphi)}\ ,\label{def:G}
\eeq
where the normalization of the functional integral is included in the 
Gaussian measure $d\mu_{p_{\Lambda,\Lambda_0}}\ $,
the covariance $p_{\Lambda,\Lambda_0}$ of $d\mu_{p_{\Lambda,\Lambda_0}}$
is given by \eqref{prop-UV-IR}, $\lambda$ is the coupling constant and $V^{(\Lambda_0)}$ is the interaction.
Heuristically speaking,
the 'degrees of freedom in the region $\Lambda^2\prec p^2\prec\Lambda_0^2$ are integrated out'.
Graphically $G_{\Lambda,\Lambda_0}(\varphi)$ is the sum of all connected Feynman diagrams with vertices
$\lambda V^{(\Lambda_0)}$, internal lines symbolizing $p_{\Lambda,\Lambda_0}$ and 
external lines symbolizing the field $\varphi$.

The interaction $V^{(\Lambda_0)}$ is usually local and
depends on $\Lambda_0$ since it is normally ordered with respect to 
$p_{0,\Lambda_0}$ (or $p_{\Lambda,\Lambda_0}$) and because it contains $\Lambda_0$-dependent local counterterms
as explained in \eqref{addct}, \eqref{addct1} below.

A main difference to our formalism is that there the interaction $V$ 
has compact support (see \eqref{interaction}); but this does not hold here,
e.g.~for the $\varphi^n$-model the unrenormalized interaction (i.e.~without counterterms) reads
$$
\int d^dx\,\Omega_{p}((\varphi(x))^n)\ , \qbox{where $\Omega_{p}(...)$
denotes normal ordering w.r.t.~$p\ $.}
$$ 
Therefore, in our formalism IR-divergences do not occur; 
but here they can appear and, hence, in general it is necessary to
introduce the IR-cutoff $\Lambda>0$. Purely massive models are an exception: they are
IR-finite also in the usual formalism and, hence, one can set $\Lambda=0$.

Computing $\frac{\d}{\d\Lambda}$ of the functional integral \eqref{def:G} one derives the flow equation.
Let $G^{(r)}_{\Lambda,\Lambda_0}$ be that term of $G_{\Lambda,\Lambda_0}$ which is of order $r$ in the coupling 
constant $\lambda$ (or equivalently in $V$).
Proceeding by induction on $r$, the flow equation expresses 
$\frac{\d G^{(r)}_{\Lambda,\Lambda_0}}{\d\Lambda}$ in terms of lower order terms 
$G^{(k)}_{\Lambda,\Lambda_0}\ ,\,\,k<r,$ which are inductively known.
Solving the flow equation
$$
G^{(r)}_{\Lambda,\Lambda_0}=G^{(r)}_{\Lambda_0,\Lambda_0}-\int_\Lambda^{\Lambda_0}d\Lambda'\,
\frac{\d G^{(r)}_{\Lambda',\Lambda_0}}{\d\Lambda'}
$$
(where $\frac{\d G^{(r)}_{\Lambda',\Lambda_0}}{\d\Lambda'}$ is expressed 
in terms of inductively known terms by the flow equation)
there appears the crucial question how to choose 
the boundary value $G_{\Lambda_0,\Lambda_0}$. Choosing for $G_{\Lambda_0,\Lambda_0}$
the unrenormalized (normally ordered) interaction $-\lambda\,\Omega_{p}(V)$,
the limit $\lim_{\Lambda_0\to\infty} G_{\Lambda,\Lambda_0}$ does not exist in general (due to the usual UV-divergences).
Therefore, one adds $\Lambda_0$-dependent local counterterms,
\beq
G_{\Lambda_0,\Lambda_0}=-\lambda\,V^{(\Lambda_0)}=-\lambda
\Omega_{p}(V)+\Lambda_0\hbox{-dependent local counterterms}\ ,\label{addct}
\eeq
such that this limit exists. The theory is 'perturbatively renormalizable' if this is possible
by a {\it finite} number of counterterms (each counterterm may be a formal power series in $\lambda$). 
In case of the $\phi^4$-interaction in $d=4$ dimensions one has to add three counterterms of the form
\beq
G_{\Lambda_0,\Lambda_0}=\lambda \Omega_{p}(\phi^4) (-1+\sum_{r\geq 1}c_{\Lambda_0,r}\lambda^r)
+\Omega_{p}(\phi^2) \sum_{r\geq 2}
a_{\Lambda_0,r}\lambda^r+ \Omega_{p}((\d\phi)^2)\sum_{r\geq 2}b_{\Lambda_0,r}\lambda^r\ ,\label{addct1}
\eeq
where $a_{\Lambda_0,r},\,b_{\Lambda_0,r},\,c_{\Lambda_0,r}$ are $\Lambda_0$-dependent numbers.

We now compare with our formalism.
\begin{itemize}
\item {\bf $\tau_{\Lambda,\Lambda_0}$ and ${\bf S}_{\Lambda,\Lambda_0}$ as functional integrals:} 
for $F\in {\cal F}$
\beq\label{FI1}
(\tau_{\Lambda,\Lambda_0}F)(\varphi) \qbox{corresponds to} 
\int d\mu_{p_{\Lambda,\Lambda_0}}(\phi)\,F(\phi+\varphi)\ ,
\eeq
since both expressions are the sum over all possible contractions of $\varphi$ in $F(\varphi)$
with propagator $p_{\Lambda,\Lambda_0}$. 

Moreover let  ${\bf S}_{\Lambda,\Lambda_0}$  be the regularized 
$S$-matrix with propagator $p_{\Lambda,\Lambda_0}\ $. 
Then, for $V\in {\cal F}$, 
\beq\label{FI2}
S_{\Lambda,\Lambda_0}(\lambda V)(\varphi) \qbox{corresponds to}  
\int d\mu_{p_{\Lambda,\Lambda_0}}(\phi)\,e^{-\lambda\,\Omega_{p_{\Lambda,\Lambda_0}}(V(\phi+\varphi))}\ , 
\eeq
since for both expressions the term $\sim\lambda^n$ is the sum over all contractions 
(with propagator $p_{\Lambda,\Lambda_0}$) between $n$ vertices, each vertex given by $V$.
Note that in the functional integral selfcontractions 
of a vertex (i.e.~tadpoles) drop out due to the normal ordering
of $V$ w.r.t.~$p_{\Lambda,\Lambda_0}$.  

Even for $F,V\in {\cal F}_{\rm loc}$ all expressions in \eqref{FI1} and \eqref{FI2} are well-defined
(i.e.~renormalization is not needed at this level) since $p_{\Lambda,\Lambda_0}\in {\cal S}(\RR^d)$.
\item {\bf Effective potential:}
Our effective potential 
\beq\label{correffpot}
V_\Lambda:={\bf S}_\Lambda^{-1}\circ {\bf S}(V) \qbox{corresponds roughly to} 
G_{\Lambda,\infty}:=\lim_{\Lambda_0\to\infty} G_{\Lambda,\Lambda_0}\ .
\eeq
For $\Lambda=0$ these expressions agree: namely in our formalism we have the value 
\beq\label{effpotL=0}
e^{V_0(\varphi)}={\bf S}(V)(\varphi)
\eeq
(see \eqref{La=0}), which is a main justification to interprete $V_\Lambda$ as effective potential. 
On the other side, in an IR-finite model, the functional integral
\beq
\lim_{\Lambda_0\to\infty} e^{G_{0,\Lambda_0}(\varphi)}=\lim_{\Lambda_0\to\infty}
\int d\mu_{p_{0,\Lambda_0}}(\phi)\,e^{-\lambda V^{(\Lambda_0)}(\phi+\varphi)}\label{G->S}
\eeq
gives also ${\bf S}(V)(\varphi)\ $.


As mentioned above, the existence of  $\lim_{\Lambda_0\to\infty} G_{\Lambda,\Lambda_0}$ involves renormalization that 
is the addition of suitable local counterterms. Also the definition 
\eqref{def:effpot} of $V_\Lambda$ presupposes renormalization, since 
$V_\Lambda$ is defined in terms of the renormalized $S$-matrix. 

\item{\bf UV-finite models:} in an UV-finite theory (e.g.~interaction $\phi^n$ in $d=2$ dimensions or $\phi^2$ for $d=3$)
$V^{(\Lambda_0)}$ \eqref{addct} depends on $\Lambda_0$ only by normal ordering. If the latter is done with respect to
$p_{\Lambda,\Lambda_0}$, we see from \eqref{def:G} and \eqref{FI2} that 
\beq
S_{\Lambda,\Lambda_0}(\lambda V)(\varphi) \qbox{corresponds to} e^{G_{\Lambda,\Lambda_0}(\varphi)}\ .
\eeq
Taking also 
\beq
{\bf S}(F)=\lim_{\Lambda_0\to\infty} {\bf S}_{\Lambda_0}(F)\ ,\quad\forall F\in {\cal F}_{\rm loc}\ ,
\eeq
(see \eqref{SL->S} below) and \eqref{SLL0} into account and choosing $\lambda=1$ we obtain
\begin{align}
G_{\Lambda,\infty}&\simeq\lim_{\Lambda_0\to\infty}\,\log\circ {\bf S}_{\Lambda,\Lambda_0}(V)\notag\\
&=\lim_{\Lambda_0\to\infty}\,\log\circ \tau_\Lambda^{-1}\circ {\bf S}_{\Lambda_0}\circ \tau_\Lambda (V)=
\log\circ \tau_\Lambda^{-1}\circ {\bf S}\circ \tau_\Lambda (V)\ .\label{flowfinite}
\end{align}
Usually this differs from
\beq
V_\Lambda={\bf S}_\Lambda^{-1}\circ {\bf S}(V)=\tau_\Lambda \circ 
\log\circ \tau_\Lambda^{-1}\circ {\bf S} (V)\ ;\label{flow1}
\eeq
an exception is $\Lambda=0$ (as generally noted in \eqref{effpotL=0}, \eqref{G->S}). 
Or, arguing somewhat differently: the difference between $V_\Lambda$ \eqref{flow1} and $G_{\Lambda,\infty}$ \eqref{flowfinite}
amounts to the difference between ${\bf S}_\Lambda^{-1}\circ {\bf S}_{\Lambda_0}$
and $\log\circ {\bf S}_{\Lambda,\Lambda_0}$ for $\Lambda_0\to\infty$, which is given by \eqref{rel-flow-op}.
The maps $V\mapsto V_\Lambda$ \eqref{flow1} and $V\mapsto G_{\Lambda,\infty}$ \eqref{flowfinite} agree
up to a similarity transformation by $\tau_\Lambda$, which is a matter of convention.

\end{itemize} 
{\bf Relation of Wilson's RG to the St\"uckelberg-Petermann RG (heuristic treatment).}
The difference in the conventions for $V_\Lambda$ and $G_{\Lambda,\infty}$ 
(\eqref{flow1} versus \eqref{flowfinite}) is no obstacle to interprete  
$V_\Lambda$ as an effective potential and, hence, to interprete the corresponding flow operators
\beq\label{WsRG}
\{{\bf S}_\Lambda^{-1}\circ {\bf S}_{\Lambda_0}\,|\,0\leq\Lambda\leq\Lambda_0<\infty\}
\eeq
{\it as a version of Wilson's RG}. Proceeding heuristically, we are now going to show that the 
restriction of the operators \eqref{WsRG} to ${\cal F}_{\rm loc}$
can be approximated by a subfamily of the St\"uckelberg-Petermann RG, for $\Lambda,\,\Lambda_0$ big enough.

For this purpose we use that, for a renormalizable model, the
limit $\Lambda\to\infty$ of ${\bf S}_\Lambda$ exists if one adds suitable {\it local} counterterms
(analogously to \eqref{addct}, \eqref{addct1}).
This addition of local counterterms can be described by $V\rightarrow Z_\Lambda(V)$ 
with an element $Z_\Lambda$ of the St\"uckelberg-Petermann group. In detail \cite{BDF}:
for all $0<\Lambda<\infty$ there exists a $Z_\Lambda\in {\cal R}_0$ with
\begin{equation}\label{SL->S}
\lim_{\Lambda\to\infty} {\bf S}_\Lambda\circ Z_\Lambda = {\bf S}\ .
\end{equation}
By ${\cal R}_0$ we mean the version of the St\"uckelberg-Petermann RG which is defined by
requiring only the conditions Starting element, Locality and Field Independence. Since 
usually ${\bf S}_\Lambda$ does not satisfy Poincar\'e invariance, almost homogeneous 
Scaling and Unitarity (and possibly violates also Smoothness in $m\geq 0$), we may not expect
that $Z_\Lambda$ fulfils the corresponding conditions.

With \eqref{SL->S} we have $\lim_{\Lambda\to\infty}{\bf S}_\Lambda\circ Z_\Lambda={\bf S}=
\lim_{\Lambda_0\to\infty} {\bf S}_{\Lambda_0}\circ Z_{\Lambda_0}\ $, from which we conclude that  
\begin{equation}
{\bf S}_\Lambda^{-1}\circ {\bf S}_{\Lambda_0}\vert_{{\cal F}_{\rm loc}}
\approx Z_\Lambda\circ Z_{\Lambda_0}^{-1}\in{\cal R}_0
\quad {\rm for} \quad \Lambda,\Lambda_0\to\infty\ .
\end{equation}
That is, for $\Lambda,\,\Lambda_0$ big enough, the restriction of the flow operators 
${\bf S}_\Lambda^{-1}\circ {\bf S}_{\Lambda_0}$ to ${\cal F}_{\rm loc}$
can be approximated by the $2$-parametric subfamily $Z_\Lambda\circ Z_{\Lambda_0}^{-1}$ 
of the St\"uckelberg-Petermann group ${\cal R}_0$.

\vskip5mm

\noindent{\bf Acknowledgments.}
The author was supported by the Deutsche Forschungsgemeinschaft through the
Institutional Strategy of the University of G\"ottingen.

This article is based on reference \cite{BDF} -- a joint work with Romeo Brunetti
and Klaus Fredenhagen. Many discussions with Romeo and Klaus have been very helpful.


\bibliographystyle{my-h-elsevier}

\end{document}